\documentclass[aps,prl,twocolumn]{revtex4}
\usepackage{amsmath}
\usepackage{graphicx}
\usepackage{multirow}
\usepackage{array}
\usepackage{inputenc}
\newcommand{\hoti}{Ho$_2$Ti$_2$O$_7$}
\newcommand{\dyti}{Dy$_2$Ti$_2$O$_7$}
\newcommand{\tbti}{Tb$_2$Ti$_2$O$_7$}
\newcommand{\tbsn}{Tb$_2$Sn$_2$O$_7$}

\newcommand{\tb}{Tb$^{3+}$}

\begin{document}
\author{Sylvain Petit$^{1}$, Pierre Bonville$^{2}$, Julien Robert$^{1}$, Claudia Decorse$^{3}$ and Isabelle Mirebeau$^{1}$}
\affiliation{$^1$ CEA, Centre de Saclay, DSM/IRAMIS/ Laboratoire L\'eon Brillouin, F-91191 Gif-sur-Yvette, France}
\affiliation{$^2$ CEA, Centre de Saclay, DSM/IRAMIS/ Service de Physique de l'Etat Condens\'e, F-91191 Gif-Sur-Yvette, France}
\affiliation{$^3$ ICMMO, Universit\'e Paris-Sud, F-91400 Orsay France}
\title{Spin liquid correlations, anisotropic exchange and symmetry breaking in \tbti}
\date{\today}
\begin{abstract}
We have studied the low energy spin dynamics between 4.6\,K and 0.07\,K in a \tbti\ single crystal sample,
by means of inelastic neutron scattering experiments. The spectra consist in a dual response, with a
static and an inelastic contribution, showing striking Q-dependences. We propose an interpretation involving
an anisotropic exchange interaction in combination with a breaking of the threefold symmetry at the rare
earth site. Simulations of the Q-dependent scattering in the Random Phase Approximation account well for
the inelastic response.
\end{abstract}
\pacs{81.05.Bx,81.30.Hd,81.30.Bx, 28.20.Cz}
\maketitle

Spin liquids now attract considerable attention in modern condensed matter physics \cite{lhuillier,balents}.
In a classical picture, the localized magnetic moments in such cooperative paramagnets keep fluctuating in a
correlated manner, failing to develop long range order down to very low temperature. From a quantum
point of view, a spin liquid ground state can be described by entangled spin wavefunctions and supports
exotic fractionalized excitations also called spinons. Geometrically frustrated magnets are
good candidates in the pursuit of such disordered quantum ground states \cite{revgingras} and one of the
celebrated examples is the \tbti\ pyrochlore. It  remains in a spin liquid state, with short range correlated
fluctuating moments, down to a temperature as low as 20\,mK \cite{gardner99,gardner01}. Since 1999, it
has been the subject of many theoretical as well as experimental works, and the origin of its spin liquid
ground state is still puzzling.

\tbti\ belongs to the same family as the \hoti\ and \dyti\ spin ices, characterized by an Ising
anisotropy along local $<111>$ axes \cite{cao}. However, the \tb\
crystal electric field (CEF) with trigonal symmetry \cite{gingras00,mirb07} has a much lower energy gap
between the ground state doublet and the first excited doublet than in spin ices (18\,K instead of 200
to 300\,K). It was suggested that, unlike in spin ices, this gap is small enough to allow admixture of
excited crystal field states, that produces an effective ferromagnetic contribution which competes with
the original antiferromagnetic interactions, and that moves \tbti\ towards a "quantum spin-ice" regime
\cite{molavian07,hertog}. More recent general descriptions of pyrochlores introduce a minimal Hamiltonian,
based on symmetry grounds, for pseudospins 1/2 (the subspace spanned by the ground doublet states
$\vert \psi_{\pm} \rangle$) \cite{shannon,balents2,savary}. These models involve an Ising exchange
constant $J_{zz}$ responsible for the spin-ice behavior, as well as three "quantum" terms $J_{\pm}$,
$J_{z\pm}$ and $J_{\pm\pm}$ that lift the macroscopic degeneracy of the spin-ice manifold and stabilize
a so called Coulomb phase or U(1) spin liquid phase, describable by an emergent U(1) gauge field. For
large quantum couplings, conventional phases are stabilized against the spin liquid. Interestingly,
in the particular case of non-Kramers ions (like \tb), these phases are characterized by ordering of
the $4f$ quadrupoles \cite{sungbin,onoda}, breaking spontaneously the threefold symmetry of the
crystal field.

Recently, we proposed a somehow more phenomenological route to point out the relevance of such a
symmetry breaking. Indeed, inelastic neutron scattering experiments have evidenced low energy spin
fluctuations \cite{mirb07,rule09,yasui}, which, because of general properties of non-Kramers ions,
cannot be explained in the nominal trigonal CEF, but can be quite naturally accounted for assuming a
breaking of the threefold symmetry \cite{bonvicfcm,bonv11,bonv09}. The same conclusion holds
in the case of the ordered spin ice parent compound \tbsn\ \cite{mirb05}, where similar, although
better defined, strong low energy fluctuations are also reported \cite{rule07,rule09,petit}.
Experimentally, such a symmetry lowering could be due to a tetragonal distortion precursor
to a $T\simeq 0$ Jahn-Teller transition. In \tbti, this is supported by thermodynamic measurements
\cite{chapuis,yaouanc}, Raman scattering \cite{lummen}, by some studies of the thermal evolution
of the elastic constants \cite{mams,nakanishi,malkin}, as well as X-ray measurements \cite{ruff07},
but the existence of this distortion is still debated in literature \cite{goto,gaulin}. Recently,
motivated by our work \cite{bonv11}, there have been several attempts to determine the characteristics
of the low energy spin dynamics and especially decide if the fluctuations are quasi-elastic or
inelastic \cite{gaulin,takatsu}, but no consensus was obtained.

In this work, we report new high accuracy and high resolution INS experiments performed on a \tbti\ single
crystal down to 0.07\,K. We unambiguously observe the controversial inelastic excitation. More precisely, we
found that the low energy response is dual, i.e. it is the sum of a rather strong elastic signal and of an
overdamped inelastic one with striking Q-dependences. By modeling the dynamic susceptibility in the
Random Phase Approximation (RPA) and using the anisotropic exchange tensor determined previously
\cite{bonv11}, we find that the hypothesis of a breaking of the threefold symmetry reproduces
the Q-dependence of the inelastic response satisfactorily.


\begin{figure}[ht]
\centerline{\includegraphics[width=8cm]{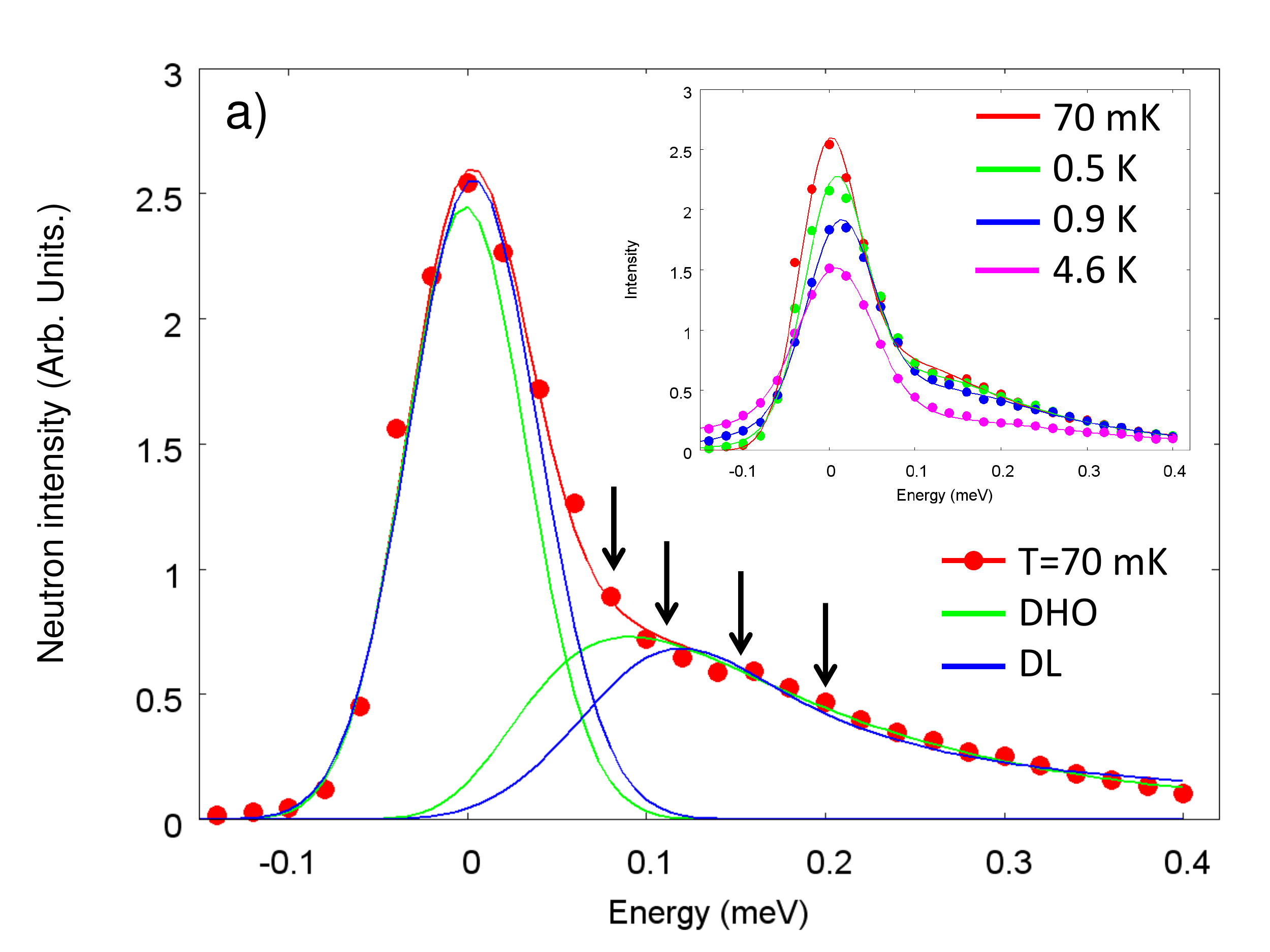}}
\centerline{\includegraphics[width=9cm]{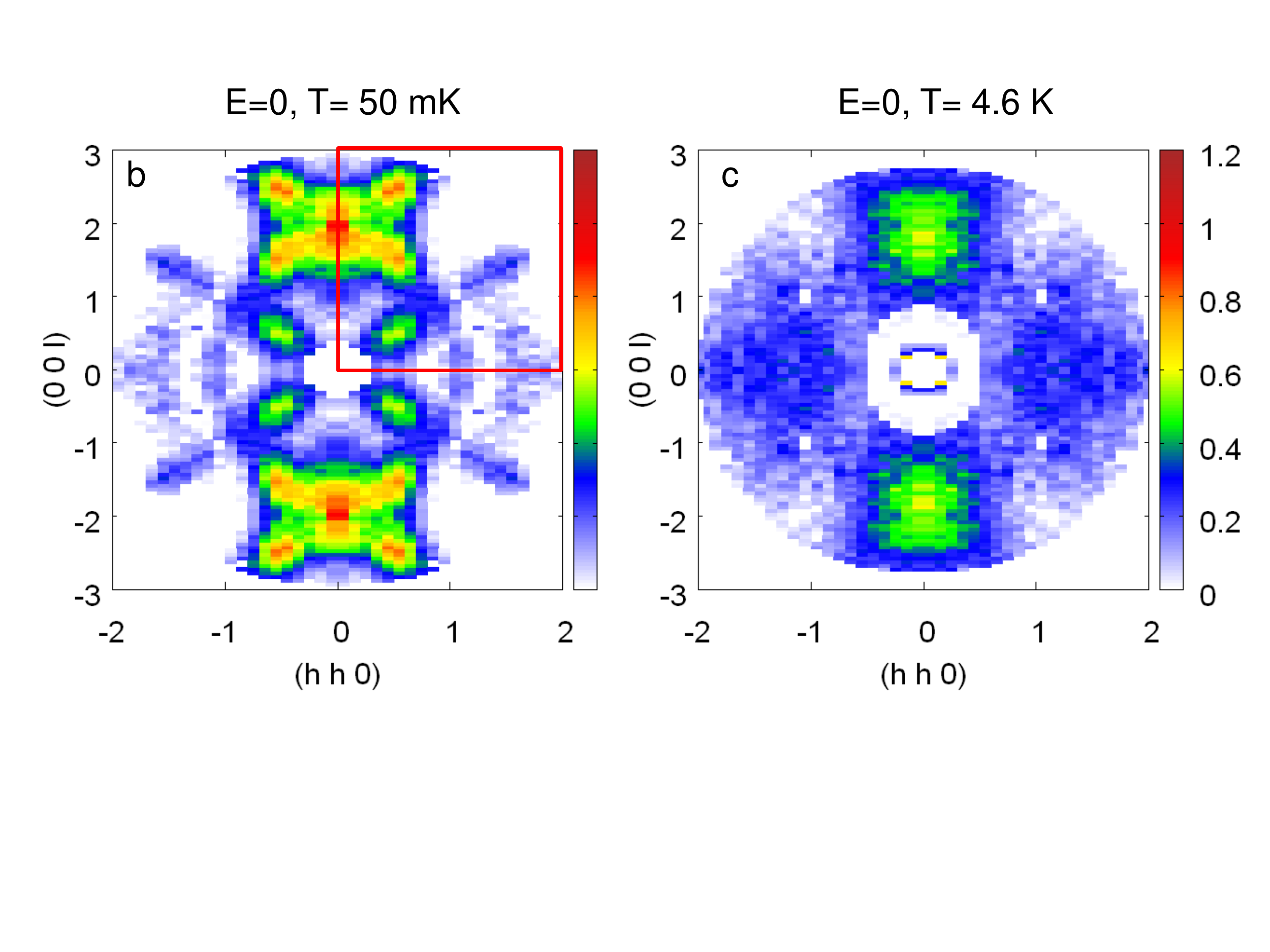}}
\vspace{-2cm}
\caption{
(color on line) {\bf a}: Neutron intensity (normalized to monitor) as a function of energy transfer
at T=0.07\,K for Q=(002). The data were recorded with a final wavevector k$_f$=1.2\AA$^{-1}$,
yielding an energy resolution (FWHM) $\Delta_0=$ 0.07\,meV. The lines show the elastic and
inelastic contributions; the red one shows the total fitted intensity in the case of a damped
harmonic oscillator (DHO) for the inelastic contribution (see text). Insert: thermal evolution of the
low energy scattering. {\bf b} and {\bf c}: Elastic magnetic scattering maps taken respectively
at 0.07\,K and 4\,K. A Q-independent background has been subtracted from the raw data in
order to remove the incoherent scattering. It was estimated from the temperature dependence
of the total elastic scattering. The color scale is identical for the two temperatures. The red box
corresponds to the points actually measured, the remaining has been deduced by symmetry.
}
\label{fig1}
\end{figure}

INS experiments have been performed on the triple axis spectrometer 4F2, installed on the cold 
neutron source at the LLB-Orph\'ee (Saclay, France) neutron facility (see the supplemental material 
for detailed information). Four \tbti\
single crystals, of total mass 11\,g, synthesized by the floating zone method, were co-aligned in the
$(hhl)$ scattering plane. We collected a series of energy scans at various wavevectors along the
high symmetry directions $(hh0)$, $(00l)$ and $(hhh)$ for different temperatures ranging from
0.07\,K up to 4\,K. We also mapped out the intensity at different constant energy transfers as a
function of wavevector in the $(hhl)$ plane. Our data provide compelling evidence
for a strong low energy response, well below the first CEF transition (at about 1.5\,meV). The high
energy resolution $\Delta_0$=0.07\,meV (full width at half maximum FWHM) allowed us to separate
an elastic contribution $I_0(Q)$ from a broad inelastic contribution $I_1(Q,\omega)$, with a
maximum around 0.15\,meV.
This is illustrated in Fig.\ref{fig1}, showing the neutron intensity at Q=(002) in the low energy range
$-0.1 \leq \omega \leq$ 0.4\,meV. The two contributions are well separated at 0.07\,K, where a
clear shoulder is observed at finite energy transfer.

\begin{figure}[ht]
\centerline{\includegraphics[width=10cm]{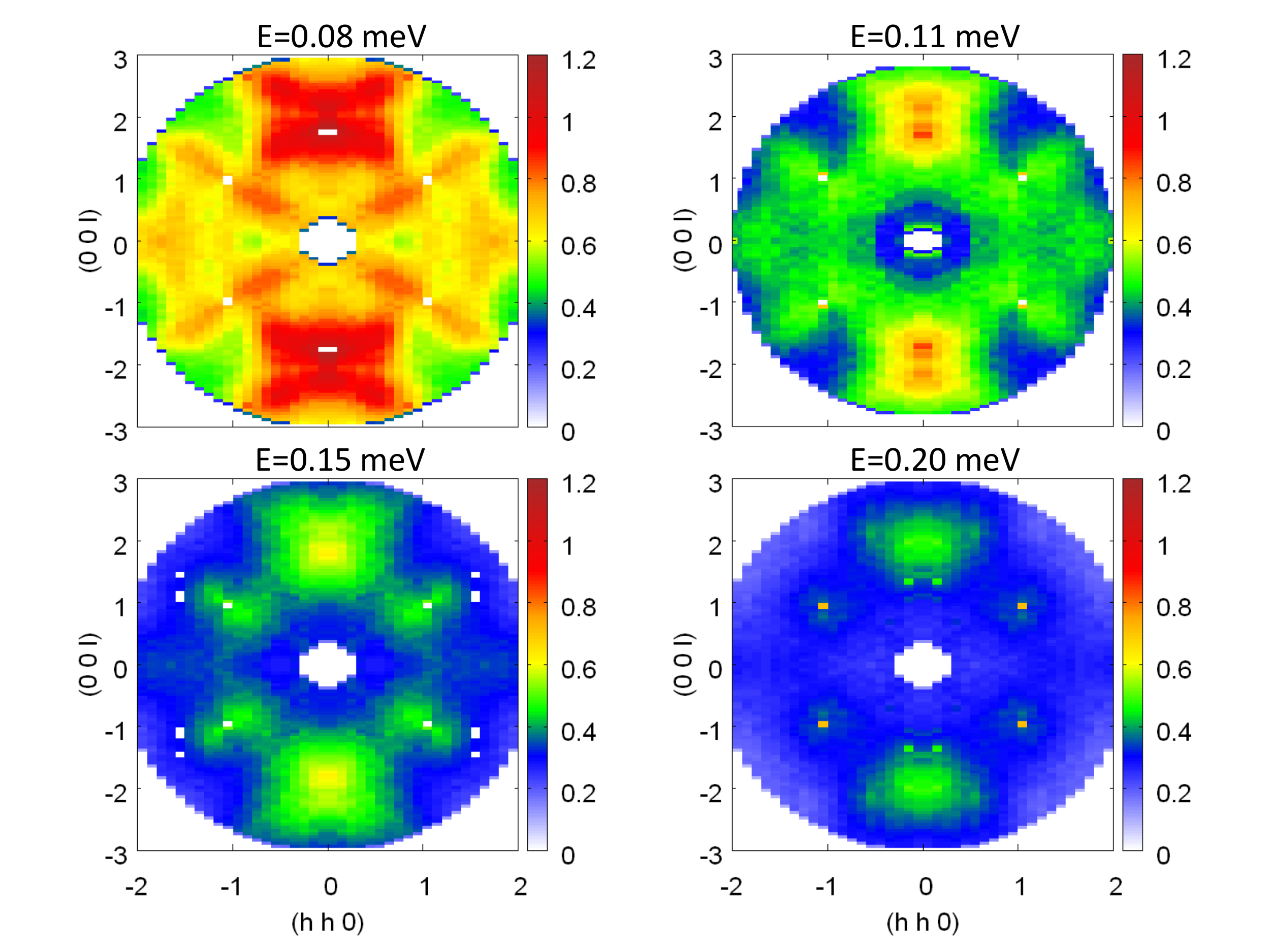}}
\caption{(color on line) Inelastic scattering maps at 0.07\,K for an energy transfer: {\bf a}: 0.08\,meV ,
{\bf b}: 0.11\,meV, {\bf c}: 0.15\,meV and {\bf d}: 0.20\,meV. These energies are indicated by
arrows in figure \ref{fig1}a.}
\label{fig2}
\end{figure}

This dual response was modeled by $I(Q,\omega) = I_0(Q) + I_1(Q,\omega)$. Note that because of the
experimental resolution, $I_0(Q)$ either corresponds to pure static or to slow fluctuations with
characteristic frequency lower than
$\nu \sim \frac{\Delta_0}{h}  \simeq$ 4\,GHz. Like in frustrated ferromagnets, different forms have been
used for $I_1(Q,\omega)$: the double Lorentzian (DL) profiles and the damped harmonic oscillator (DHO),
with characteristic energy $\omega_1$ and damping (HWHM) $\Gamma_1$ \cite{hennion}. After convolution
with the experimental resolution function, we found that the DHO profile yields the best agreement with the
data. Fitting the energy scans recorded along the high symmetry direction shows almost no change of
$\omega_1$ in Q-space: $\omega_1 \approx$ 0.20\,meV,
while the damping $\Gamma_1$ is large and has the same magnitude: $\Gamma_1 \approx \omega_1$.
Maps of $I_0(Q)$ measured at 0.07\,K and 4\,K, shown in Fig.\ref{fig1}, reveal
a ground state with strongly anisotropic short range correlations. At 0.07\,K, we observe a complex pattern
resembling an array of "butterflies" pinned at (002) and, with a smaller intensity, at (220)-type positions.
Superimposed on that structure, the map shows lobes centered around ($1/2$, $1/2$, $1/2$)-type positions,
elongated along the $(hhh)$ directions, and separated by a pinch-point at (111). A series of inelastic Q-space
maps, taken at 0.07\,K at different energy transfers, are shown in Fig.\ref{fig2}. As the energy transfer
increases, the pinch-point at Q=(111) is progressively "filled" while the intensity at (1/2,1/2,1/2)-type
positions decreases. With increasing temperature up to 4\,K, the intensity of the two signals progressively
weakens and the inelastic signal progressively merges into the elastic one. The lobes centered at
(1/2,1/2,1/2) persist up to 0.5\,K, but become hardly visible at 4\,K, with only weak maxima at (002) and
(220)-type positions.

We proceed now to the analysis of the inelastic maps in the $(hhl)$ plane of the reciprocal space.
As mentioned in the introduction, the hypothesis of a breaking of the trigonal symmetry, possibly
due to a precursor Jahn-Teller distortion developing at low temperature, has been put forward to
explain the peculiar properties of \tbti\ \cite{bonvicfcm,bonv11}. The relevant CEF Hamiltonian
writes:
\begin{equation}
{\cal H_{\rm CEF}} = {\cal H}_{\rm trig} + \frac{D_Q}{3} [ 2J_x^2+J_z^2+\sqrt{2} (J_xJ_z+J_zJ_x)],
\label{fCEF}
\end{equation}
where ${\cal H}_{\rm trig}$ represents the nominal trigonal crystal field and the second term a
small tetragonal distortion along a cubic [001] axis ($\vec J$ is the total angular momentum; for
\tb, J=6 and the Land\'e factor is $g_J$ = 3/2).
The breaking of the threefold symmetry results in a degeneracy lifting and in a mixing of the two states of the ground
doublet $\vert \psi_{\pm} \rangle$, and stabilizes a CEF singlet state described by the symmetric
wavefunction $\vert \psi_s \rangle = \frac{1}{\sqrt{2}}[\vert \psi_+ \rangle + \vert \psi_- \rangle]$
\cite{bonvicfcm,bonv11}. The first excited state is then the antisymmetric combination:
$\vert \psi_a \rangle = \frac{1}{\sqrt{2}}[\vert \psi_+ \rangle - \vert \psi_- \rangle]$, separated
from $\vert \psi_s \rangle$ by a quantity $\delta$ proportional to the amplitude of the distortion.
In the neutron spectra, this gives rise to an excitation at an energy close to $\delta$, unveiling the
transition from $\vert \psi_s \rangle$ to $\vert \psi_a \rangle$. Its intensity is
proportional to $\sum_{a=x,y,z} |\langle \psi_{s} \vert J^a \vert \psi_{a} \rangle|^2$. A
straightforward calculation shows that, whereas the general properties of non-Kramers ions impose
$\langle \psi_+  \vert \vec{J} \vert \psi_- \rangle \equiv 0$, the entanglement of 
$\vert \psi_{\pm} \rangle$ yields a sizeable $\langle \psi_a  \vert \vec{J} \vert \psi_s \rangle$ 
matrix element and thus a large cross section. This accounts for the presence of the low energy 
inelastic line in Fig.\ref{fig1} {\bf a}. In the absence of entanglement, its intensity would be
vanishingly small \cite{petit}.

We then introduce the exchange/dipolar Hamiltonian
widely accepted for pyrochlores \cite{gingras00}, with an anisotropic exchange tensor
${\cal \tilde J}_{ex}$ taken to be diagonal in the $(\vec u,\vec v,\vec w)$ frame linked with a Tb-Tb
bond along $\vec w$ \cite{malkin1}. For instance, for the bond along [110], this frame is defined by:
$\vec u$=(0,0,1), $\vec v=\frac{1}{\sqrt{2}}$(1,-1,0) and $\vec w=\frac{1}{\sqrt{2}}$(1,1,0). The
link between this exchange tensor and that derived in Ref.\cite{savary} is discussed in the Supplemental
material. For not too large antiferromagnetic ${\cal J}_{ex}$, depending on the value of the 
${\cal \tilde J}_{ex}/\delta$ ratio, the mean field ground state of the model is either an ordered 
spin ice phase or a singlet yielding no magnetic order, which is quite similar to Bleaney's result for 
non-Kramers rare earth ions with a singlet CEF ground state \cite{bleaney}.

The set of parameters
($D_Q$ and the exchange tensor components) that stabilizes $\vert \psi_s\rangle$ as the ground state
has been determined previously \cite{bonv11}: $D_Q$=0.25\,K, ${\cal J}_u=-0.07$\,K, ${\cal J}_v=-0.2$\,K
and ${\cal J}_w=-0.1$\,K (a negative sign for ${\cal J}_i$ means an antiferromagnetic coupling). With this
$D_Q$ value, a low energy inelastic line is expected at $ \delta \simeq \omega_1 \simeq 0.22$\,meV in the
inelastic neutron data. We then calculated the corresponding cross section in the RPA approximation
\cite{jensen,kao,maestro} as described in the Supplemental material. A finite width (HWHM)
$\Gamma_1$=0.20\,meV was introduced in the single site susceptibility for the lower transition
$\vert \psi_s \rangle \leftrightarrow \vert \psi_a \rangle$, according to the DHO fit. The widths
for the other transitions were set at a smaller value, with no influence on the result.

\begin{figure}[t]
\centerline{
\includegraphics[width=9cm]{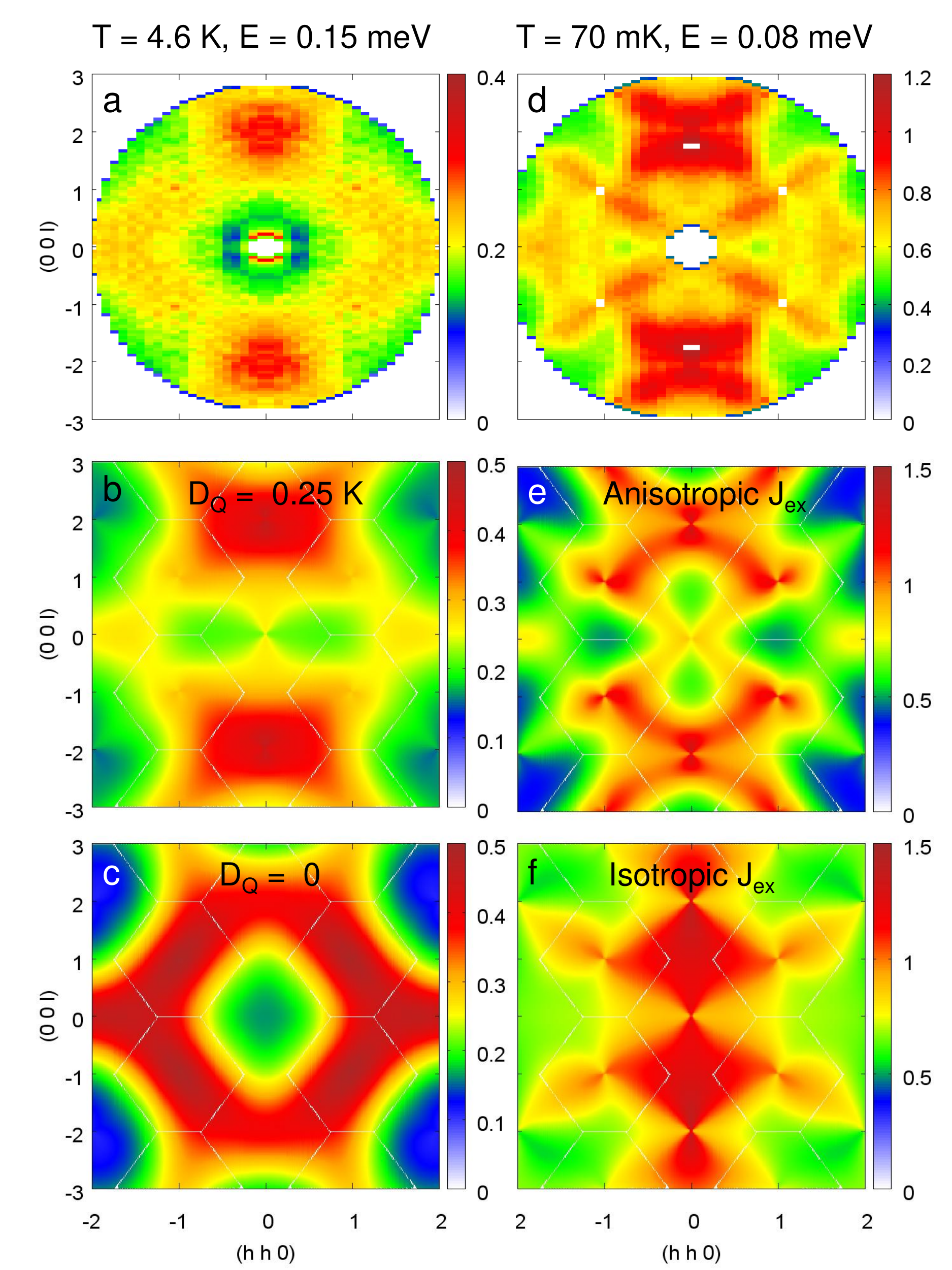}
}
\caption{(color on line) Inelastic scattering maps in \tbti.
{\bf Left column}: $T$=4.6\,K and energy transfer 0.15\,meV: {\bf a}: Experimental data; {\bf b} and {\bf c}: 
Simulations respectively with a tetragonal distortion $D_Q=0.25$\,K and without. The simulation in {\bf c} has 
been multiplied by 7000 to obtain the same intensity as in {\bf b}. {\bf Right column}: $T$=0.07\,K and energy 
transfer 0.08\,meV: {\bf d}: Experimental data; {\bf e} and {\bf f}: Simulations in the presence of a tetragonal 
distortion $D_Q=0.25$\,K, with respectively an anisotropic and an isotropic exchange tensor. From these latter 
maps, it is clear that the Q-dependence of the scattering strongly depends on the anisotropy, and that it is not 
reproduced with isotropic exchange.}
\label{fig3}
\end{figure}

Figure \ref{fig3} compares the experimental inelastic scattering \ref{fig3}{\bf a} at 4.6\,K for an
energy transfer 0.15\,meV with simulations obtained with (\ref{fig3}{\bf b}) and without
(\ref{fig3}{\bf c}) the tetragonal distortion. The simulated map with distortion is much closer to
the experimental data, reproducing the two intense spots at (002) and the rhomb-shaped lesser
intensity scattering centered at (220). The map simulated without distortion, i.e. assuming a
degenerate doublet as ground state, is quite different, but it bears resemblance with the diffuse
scattering pattern observed at 9\,K \cite{gardner01,kao}. This points to the fact that the symmetry
breaking is already present at 4\,K, in agreement with the Raman data \cite{lummen}. At the base
temperature (70 mK\,), the simulated inelastic map \ref{fig3}{\bf e}, for an energy transfer 0.08\,meV,
also reproduces correctly the structure in Q-space reproduced for convenience in \ref{fig3}{\bf d}: the
observed butterfly-shaped structures with centers at (002), the high intensity lines along (111) and
the pinch-points at (111)-type positions are present. Finally, we examine the influence of exchange
anisotropy. Figure \ref{fig3}{\bf f} represents a simulation at 0.07\,K performed in the same
conditions as \ref{fig3}{\bf e}, but for an isotropic exchange constant ${\cal J} = -$0.04\,K
\cite{bonv11}. Its structure does not show the "butterfly-shaped" pattern observed in the data.
Our calculations actually reveal that such a pattern arises only for an anisotropic exchange tensor
close to that derived for \tbti\ (${\cal J}_u=-0.07$\,K, ${\cal J}_v=-0.2$\,K, ${\cal J}_w=-0.1$\,K).

These RPA simulations thus point to the relevance of the symmetry breaking, leading to a sizeable
inelastic scattering, while the anisotropy of the exchange coupling does reproduce quite well the
Q-dependence of the inelastic response. We thus have shown that at this mean field level the
model captures important features relevant to the physics of \tbti. It is however inadequate to handle 
other complex features. Especially, the elastic component, given by the Curie term in the expression 
of the single site susceptibility (see Supplemental Material), is zero since the CEF singlets are non magnetic.
This is a limitation of the mean field approach, but we notice that experimentally, the Q dependence
of the elastic signal is similar to that of the inelastic one, suggesting that the same set of
interactions is at play.

To go beyond the present model, 
different approaches could be suggested. At the level of a single tetrahedron,
exact diagonalization of the Tb spin system, currently under way, could possibly capture part of
the elastic scattering. Coupling the moments by defects, such as a distribution of distortions for 
instance, could yield static correlations and the exotic transition seen in susceptibility and specific 
heat\cite{elsa,sarah,yaouanc}. Other approaches suggest that tunneling processes between
different spin-ice configurations also give rise to quasi-elastic scattering \cite{shannon}.
Finally, we note that recent theoretical models for non-Kramers ions predict new phases 
characterized by a spontaneous breaking of the threefold symmetry of the crystal field 
\cite{sungbin,onoda} as well as an ordering of the $4f$ quadrupoles. Since a coupling between 
lattice degrees of freedom and such quadrupoles is quite natural, this could provide a new
basis for our interpretation.

\begin{figure}[t]
\includegraphics[width=8cm]{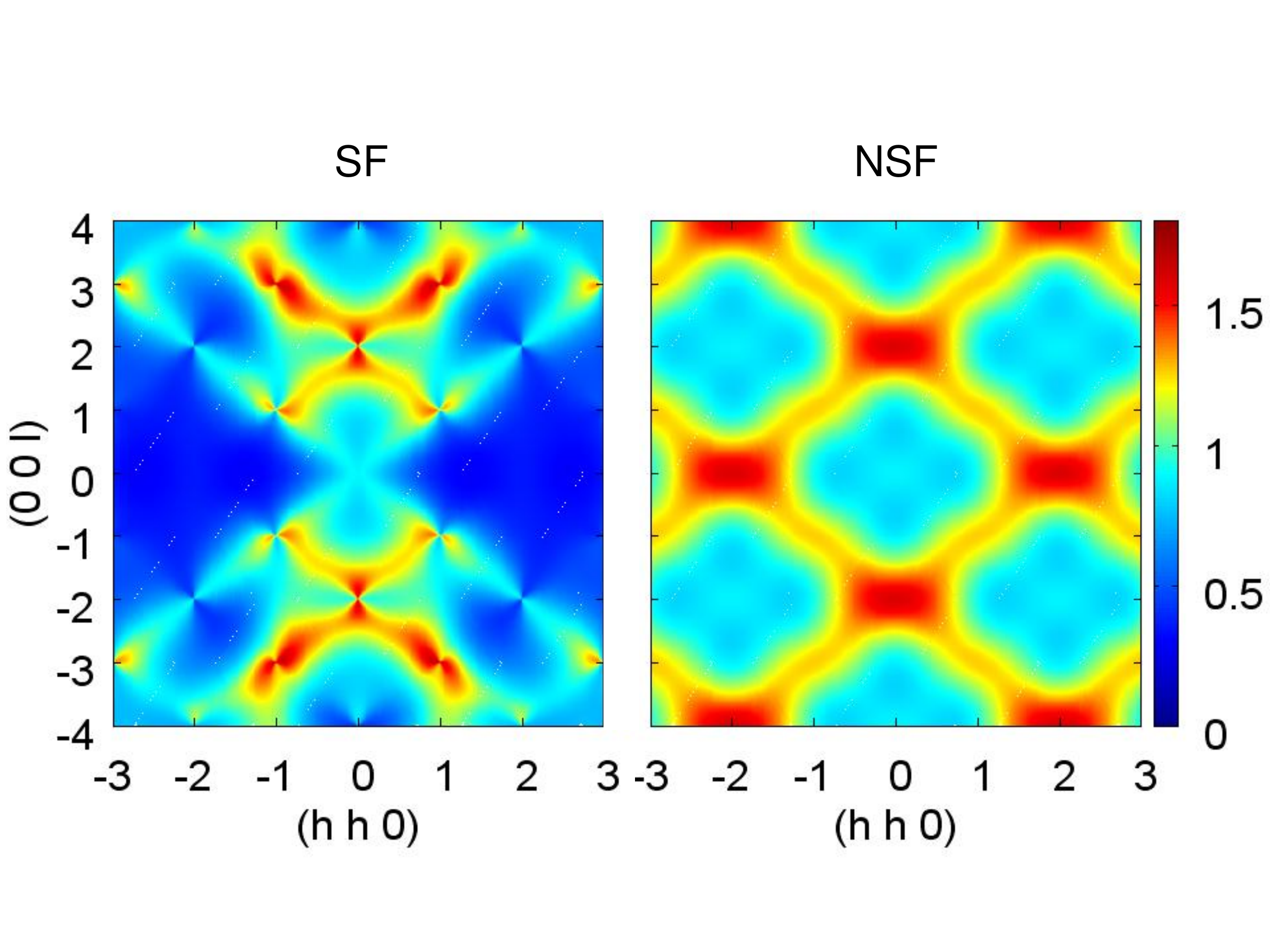}
\caption{Simulated diffuse SF (left) and NSF (right) scattering maps at 0.07\,K in \tbti.}
\label{sf-nsf}
\end{figure}

After this paper was submitted, we became aware of an independent neutron
scattering study by Fennell {\it et al} \cite{fennell1}. It was performed in different experimental
conditions, being an energy integrated experiment, whereas our measurements are performed
with energy analysis of the outcoming neutrons, thus yielding a direct access to the spin fluctuation
spectrum. Fennell {\it et al} also report on polarized neutron scattering experiments, which allow 
to distinguish the correlations of spin components along the vertical axis $[1,-1,0]$, in the Non-Spin-Flip
(NSF) channel, and correlations of spin components perpendicular to $\vec{Q}$ within the scattering 
plane in the Spin-Flip (SF) channel. Figure \ref{sf-nsf} shows, for both channels, energy-integrated 
maps at 0.07\,K calculated in the framework of  our model, that capture the experimental data 
(Fig. 2 A and F of ref.\cite{fennell1}) quite well. The hypothesis of a symmetry breaking and anisotropic
exchange are thus compulsory to understand the inelastic response, as we have shown above, but
also the diffuse scattering observed in ref. \cite{fennell1}.

In conclusion, our INS experiments in \tbti\ performed down to 0.07\,K show unconventional spin
dynamics with peculiar features in Q-space, such as pinch-points and "butterfly-shaped" patterns
in the $(hhl)$ plane, very different from those observed in classical spin ices. We propose an
interpretation in terms of a breaking of the threefold symmetry of the crystal field at the \tb\ sites.
We present calculations of the inelastic scattering and diffuse cross-section of polarized neutrons 
in the frame of the RPA approximation which support this picture and confirm the anisotropy of the 
exchange tensor derived previously for \tbti.

We would like to acknowledge fruitful discussions with P. Holdsworth, M. Gingras, B. Canals,
E. Lhotel as well as with B. Hennion.


\end{document}